\documentclass[aip,pop,reprint,groupedaddress]{revtex4-1}

\usepackage{graphicx,color}
\usepackage{amssymb,amsmath}
\usepackage[normalem]{ulem}

\begin{document}

\title{Improving particle beam acceleration in plasmas}

\author{M. C. de Sousa}
\email[]{meirielenso@gmail.com}
\author{I. L. Caldas}
\email[]{ibere@if.usp.br}
\affiliation{Institute of Physics, University of S\~ao Paulo, S\~ao Paulo, S\~ao Paulo, Brazil}

\date{\today}

\begin{abstract}
The dynamics of wave-particle interactions in magnetized plasmas restricts the wave amplitude to moderate values for particle beam acceleration from rest energy. We analyze how a perturbing invariant robust barrier modifies the phase space of the system and enlarges the wave amplitude interval for particle acceleration. For low values of the wave amplitude, the acceleration becomes effective for particles with initial energy close to the rest energy. For higher values of the wave amplitude, the robust barrier controls chaos in the system and restores the acceleration process. We also determine the best position for the perturbing barrier in phase space in order to increase the final energy of the particles.
\end{abstract}

\maketitle

%---------------------------------------------------------------------------------
\section{Introduction}
\label{Sec:Introduction}

Wave-particle interactions are generally a nonlinear process \cite{Lichtenberg1992, Shukla1986, Mendonca2001, Pakter1995} that may present regular and chaotic behavior in its phase space \cite{Lichtenberg1992, Shukla1986}. The predominant behavior depends on the parameters of the system, especially the amplitude of the wave that perturbs the particles. Chaotic trajectories are useful for particle heating \cite{Karney1978}, whereas regular resonant islands are responsible for particle acceleration \cite{Tajima1979, Shukla1986, Pakter1995, Spektor2004, deSousa2010, deSousa2012}. Both types of trajectories present many applications in different areas of Physics such as plasma physics \cite{Shukla1986}, accelerator physics \cite{Davidson2001, Edwards2004, Shukla1986}, fusion science \cite{Fisch1987, Berk1992}, and astrophysics \cite{Shukla1986}.

The recent development of applications with increasingly fast particles requires the study of mechanisms to improve particle acceleration. In the resonant islands, the amount of energy transferred from the wave to the particles is proportional to the wave amplitude \cite{deSousa2010}. To increase the energy transfer, one could simply increase the wave amplitude. However, this strategy has a limit since the system becomes chaotic as we raise the wave amplitude.

Control of chaos is a challenge in several areas \cite{Cary1982, Cary1986, Shinbrot1993, Spatschek1998, Baptista1998, Gauthier2003, Hudson2004, Chandre2006, Souza2007, Sanjuan2010}, and different methods have been developed to address the problem \cite{Pontryagin1961, Ott1990, Lima1990, Shinbrot1993, Lai1993, Ott1995, Vittot2004}. For Hamiltonian systems, there are global methods that control chaos in the entire phase space \cite{Vittot2004, CiraoloPRE2004, CiraoloJPA2004, Chandre2005, deSousa2012}. These methods create invariant tori in the whole phase space that reduce the chaotic regions whereas preserving the original structure. For wave-particle interactions as the ones we investigate in this paper, the addition of multiple invariant tori to the system has been verified experimentally in Ref. \cite{Chandre2005}.

An alternative to global methods is the implementation of a single invariant barrier in phase space. In some cases, the barrier acts only locally, not affecting regions far from it in phase space \cite{Vittot2005, Ciraolo2005, Chandre2006, Macor2007}. In other methods, the barrier is created in such a way that it is able to alter the structures of phase space in regions far from it \cite{EgydiodeCarvalho2005, EgydiodeCarvalho2009, Martins2010, Martins2011}.

Some methods of control of chaos have already been used to control chaotic transport in plasmas confined by magnetic fields \cite{CiraoloPRE2004, CiraoloJPA2004, Chandre2006}, to control chaotic transport in turbulent electric fields \cite{Ciraolo2005}, and to regularize particle dynamics in wave-particle interactions \cite{Chandre2005, Macor2007, deSousa2012}. In Ref. \cite{deSousa2012}, we used a global method to control chaos in the system while preserving the main structures in phase space. Nonetheless, the global method used in Ref. \cite{deSousa2012} was not intended for particle acceleration from low initial energies.

In this paper, we present an alternative method. We analyze the alterations in phase space caused by a single perturbing invariant robust barrier with the purpose of controlling chaos and improving particle acceleration from low initial energies in plasma based accelerators. The perturbing invariant curve, generated by an external perturbation, is introduced as a robust torus \cite{EgydiodeCarvalho2009, Martins2010, Martins2011}. It means that the perturbation caused by the wave vanishes at the exact position of the perturbing robust barrier, irrespective of the wave parameters. Therefore, the robust barrier is not affected by the wave parameters, in contrast with the KAM tori, which are destroyed by increasing wave amplitudes. For continuity reasons, the perturbation caused by the wave is reduced around the robust barrier. On the other hand, the barrier amplifies the action of the wave on the particles for regions far from it in phase space.

An invariant robust barrier is generally used to separate two regions in phase space. In plasma physics, a robust barrier is used to control the transport, to improve plasma confinement, and to prevent the plasma from reaching and damaging the tokamak walls \cite{EgydiodeCarvalho2009, Martins2010, Martins2011}. In Ref. \cite{Marcus2008}, electrodes were placed at the plasma edge to modify the plasma electric field and create a robust transport barrier in tokamaks. References \cite{Frerichs2012, FraileJunior2017} also present a model that produces effects similar to a robust barrier. This model is used to describe the plasma response to a resonant magnetic perturbation generated by an external current.

Most of the papers analyze the local effects of a perturbing robust barrier, such as the regularization of phase space \cite{Martins2011}, the reduction in islands size \cite{Martins2011}, and the reorganization of resonant islands \cite{EgydiodeCarvalho2005, EgydiodeCarvalho2009, Martins2010} around the barrier. In this work, we approach the effects caused by the perturbing barrier on regions far from it in phase space. We show that depending on its position, the barrier may alter the position of periodic points, increase or reduce the islands size, and greatly improve the process of particle acceleration in plasmas.

We analyze a relativistic low density beam confined by a uniform magnetic field, and interacting with a stationary electrostatic wave that propagates perpendicularly to the magnetic field. In Ref. \cite{deSousa2010}, we showed that the main resonance of this system may present a triangular shape according to the parameters of the wave. In this scenario, the final energy of particles is maximum, whereas its initial energy is minimum and close to their rest energy. Nonetheless, we observe that this optimum condition for particle acceleration only occurs for a limited interval in the wave amplitude. When the wave amplitude is low, the initial energy of particles is far from their rest energy. For higher values of the wave amplitude, chaotic trajectories destroy the resonant islands used for particle acceleration.

To overcome these problems, we modify the phase space of the system using a perturbing invariant robust barrier. We show that, when properly located, the perturbing barrier preserves the main structures in phase space, but alters the resonant islands in such a way that improves particle acceleration. We calculate analytically the position of the barrier, as a function of the wave parameters, that makes it possible to accelerate particles from their rest energy. We also obtain the values of parameters for which the islands of the main resonance disappear from phase space. From these results, we define an interval for the allowed positions of the perturbing barrier.

We determine the best position for the perturbing barrier in phase space, and we show that it is a very efficient method to improve particle acceleration in plasmas. For both low and high wave amplitudes, it reduces the initial energy of particles to their rest energy, and increases their final energy. For high values of the wave amplitude, the perturbation controls chaos in the system and restores resonant islands used for particle acceleration. Therefore, in the perturbed system, we are able to achieve the optimum condition and regularly accelerate particles from their rest energy for a much larger interval in the wave amplitude.

The paper is organized as follows. We present the original system in Section \ref{Sec:OriginalSystem}, and we discuss its drawbacks for particle acceleration. In Section \ref{Sec:RobustTorus}, we alter the phase space of the system with a perturbing robust barrier. We calculate an interval for the allowed positions of the perturbing barrier, and we determine its best position in phase space, such that the initial energy of particles is minimum, whereas their final energy is maximum. In Section \ref{Sec:ResultsDiscussion}, we build the phase space of the perturbed system, and we discuss how it compares to the original system. For low wave amplitudes, the barrier reduces the initial energy of particles to their rest energy, and slightly increases their final energy. When the wave amplitude is high, the perturbation controls chaos and restores the process of particle acceleration. In Section \ref{Sec:Conclusions}, we present our conclusions.

%---------------------------------------------------------------------------------
\section{Original system}
\label{Sec:OriginalSystem}

We analyze a relativistic low density beam confined by an external uniform magnetic field $\mathbf{B}=B_0\hat{z}$ with vector potential $\mathbf{A}=B_0x\hat{y}$. The particles in the accelerator interact with a stationary electrostatic wave given as a series of periodic pulses propagating perpendicularly to $\mathbf{B}$ with period $T$, wave vector $\mathbf{k}=k\hat{x}$, and amplitude $\varepsilon/2$. Following Refs. \cite{deSousa2010,deSousa2012}, the dimensionless Hamiltonian that describes the dynamics transverse to the magnetic field is given by
\begin{align}
	\label{eq:HOriginalCartY}
	H = & \sqrt{1 + p_x^2 + (p_y - x)^2}  \nonumber \\*
	    & + \frac{\varepsilon}{2} \cos(kx) \displaystyle\sum_{n=-\infty}^{+\infty} \delta(t-nT) .
\end{align}

Since $\dot p_y = - \partial H / \partial y = 0$, $p_y$ is a conserved quantity. For simplicity we assume, with no loss of generality, $p_y = 0$. We point out that this condition is not a physical constraint. Although $p_y$ is conserved and we assume it to be zero, $dy/dt$ is not null and the particles describe a two-dimensional movement in the $xOy$ plane.

With the assumption that $p_y = 0$, Hamiltonian (\ref{eq:HOriginalCartY}) becomes
\begin{equation}
	\label{eq:HOriginalCart}
	H = \sqrt{1 + p_x^2 + x^2} + \frac{\varepsilon}{2} \cos(kx) \displaystyle\sum_{n=-\infty}^{+\infty} \delta(t-nT) .
\end{equation}
Working with the action-angle variables $(I,\theta)$ of the unperturbed integrable system, we rewrite Hamiltonian (\ref{eq:HOriginalCart}) as \cite{deSousa2010,deSousa2012}
\begin{equation}
	\label{eq:HOriginalITheta}
	H = \sqrt{1 + 2I} + \frac{\varepsilon}{2} \cos(k\sqrt{2I}\sin\theta) \displaystyle\sum_{n=-\infty}^{+\infty} \delta(t-nT) ,
\end{equation}
where $x = \sqrt{2I}\sin\theta$ and $p_x = \sqrt{2I}\cos\theta$.

From Hamiltonians (\ref{eq:HOriginalCart}) and (\ref{eq:HOriginalITheta}), we obtain a map that describes the time evolution of the system between two consecutive wave pulses. In Hamiltonian (\ref{eq:HOriginalCart}), the perturbative term associated with the wave is a function of the $x$ variable. Therefore, only the momentum $p_x$ undergoes an abrupt change across a wave pulse. We use this property to integrate the Hamilton's equations, and obtain the time evolution of $(x, p_x)$ across a wave pulse \cite{deSousa2010,deSousa2012}:
\begin{subequations}
	\label{eq:TEAcrossPulsexp}
	\begin{align}
		x_n^+ = & x_n , \\*
		p_{x,n}^+ = & p_{x,n} + \frac{\varepsilon k}{2} \sin(k x_n) ,
	\end{align}
\end{subequations}
where $(x_n, p_{x,n})$ are the values of $(x,p_x)$ immediately before the $n^{th}$ wave pulse centered at $t=nT$, and $(x_n^+, p_{x,n}^+)$ are the values of the variables immediately after the $n^{th}$ wave pulse.

Between two consecutive wave pulses, the system is integrable and Hamiltonian (\ref{eq:HOriginalITheta}) becomes independent of the $\theta$ variable. For this reason, we use Hamiltonian (\ref{eq:HOriginalITheta}) to calculate the exact time evolution of $(I,\theta)$ between two wave pulses \cite{deSousa2010,deSousa2012}
\begin{subequations}
	\label{eq:TEBetweenPulsesITheta}
	\begin{align}
		I_{n+1} = & I_n^+ , \\*
		\theta_{n+1} = & \theta_n^+ + \frac{T}{\sqrt{1 + 2I_n^+}} \pmod{2\pi} ,
	\end{align}
\end{subequations}
where $(I_{n+1}, \theta_{n+1})$ are the values of $(I,\theta)$ immediately before the $(n+1)^{th}$ wave pulse centered at $t=(n+1)T$.

Putting expressions (\ref{eq:TEAcrossPulsexp}) and (\ref{eq:TEBetweenPulsesITheta}) together, we obtain an exact and explicit map that describes the time evolution of the system in the $(I,\theta)$ variables \cite{deSousa2010,deSousa2012}
\begin{subequations}
	\label{eq:MapOriginal}
	\begin{align}
		I_{n+1} = & \frac{1}{2} \Biggl\{2I_n\sin^2\theta_n + \biggl[\sqrt{2I_n}\cos\theta_n \biggr. \Biggr.     \nonumber \\*
				  & + \left. \left. \frac{\varepsilon k}{2}\sin(k \sqrt{2I_n} \sin\theta_n)\right]^2\right\} , \\*
		\theta_{n+1} = & \arctan \left[\frac{2\sqrt{2I_n}\sin\theta_n}{2\sqrt{2I_n}\cos\theta_n + \varepsilon k\sin(k \sqrt{2I_n} \sin\theta_n)} \right]     \nonumber \\*
					   & + \frac{T}{\sqrt{1 + 2I_{n+1}}} \pmod{2\pi} ,
	\end{align}
\end{subequations}
with $\theta$ defined in the interval $[0,2\pi]$ (since $\theta_n^+$ from expression (\ref{eq:TEBetweenPulsesITheta}b) equals $\arctan (x_n^+/p_{x,n}^+)$, and we calculate the individual values of $(x_n^+, p_{x,n}^+)$ from expressions (\ref{eq:TEAcrossPulsexp}), we are able to determine in which quadrant $\arctan (x_n^+/p_{x,n}^+)$ lies and, consequently, define $\theta$ between $[0,2\pi]$).

Using the canonical transformation $I = (x^2 + p_x^2)/2$ and $\theta = \arctan (x/p_x)$, we can write map (\ref{eq:MapOriginal}) in the Cartesian coordinates $(x, p_x)$. However, the use of action-angle variables is more convenient because the action $I$ is conserved in the absence of the perturbation caused by the wave. Moreover, action-angle variables are appropriate to the study of resonances and the associated process of particle acceleration that we investigate in this paper.

\begin{figure*}[!tb]
	\centering
	\includegraphics{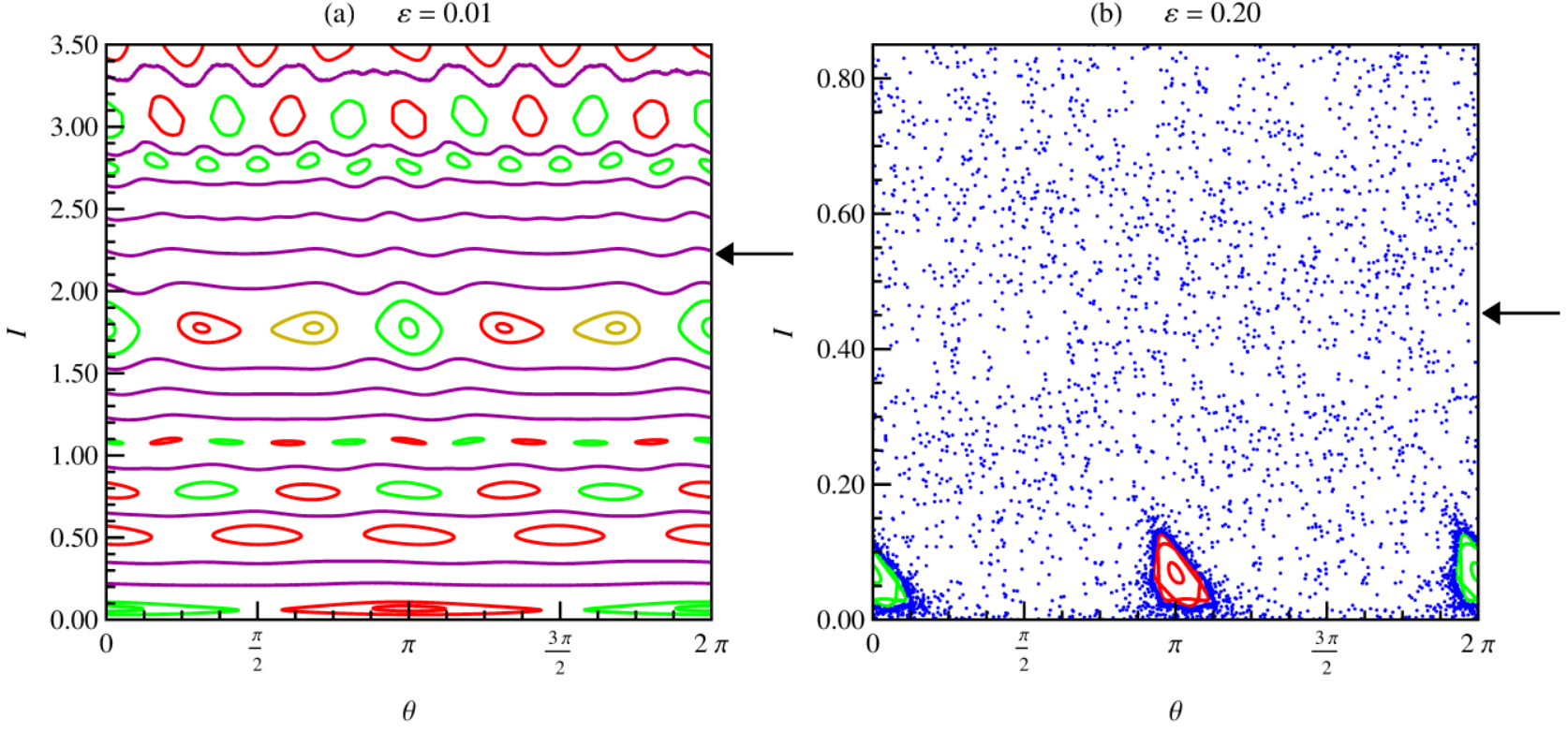}
	\caption{Phase space of the original system for $T=2 \pi (1 + 1/15)$, $k=4$, and (a) $\varepsilon=0.01$; (b) $\varepsilon=0.20$. The arrows indicate the position where the perturbing robust barrier will be placed.}
	\label{fig:OriginalSystem}
\end{figure*}

Figure \ref{fig:OriginalSystem} shows the phase space of the system built from map (\ref{eq:MapOriginal}) for different values of $\varepsilon$. In both panels, $T=2 \pi (1 + 1/15)$ and $k=4$. In Fig. \ref{fig:OriginalSystem}.(a), the wave amplitude $\varepsilon$ is small and the phase space is regular, presenting invariant tori and resonant islands. Increasing the value of $\varepsilon$ as in Fig. \ref{fig:OriginalSystem}.(b), the phase space is dominated by chaos and only a few resonant islands remain.

The system described by Hamiltonian (\ref{eq:HOriginalITheta}) presents an infinite number of resonances in different positions along its phase space \cite{deSousa2013}, as can be seen in Fig. \ref{fig:OriginalSystem}.(a). In the main resonance of the system (the (1,1) resonance), the wave frequency equals the unperturbed system frequency, i.e., each wave pulse corresponds to a full particle turn in the plane perpendicular to the magnetic field. When the wave period is slightly larger than the cyclotronic period $T_\text{c} = 2\pi$ as in Fig. \ref{fig:OriginalSystem}, the main resonance is close to the axis $I=0$.

Figure \ref{fig:OriginalSystemRes11} shows an amplification of Fig. \ref{fig:OriginalSystem} in the main resonance region. In the resonant islands, the wave transfers a great amount of energy to the particles and they are regularly accelerated \cite{Tajima1979, Shukla1986, Pakter1995, Spektor2004, deSousa2010, deSousa2012}. Since the main resonance is close to the axis $I=0$, which corresponds to the rest energy of the particles, the initial energy of particles in these islands may be very low.

In Ref. \cite{deSousa2010}, we showed that the main resonance may present a triangular shape according to the parameters of the wave. We obtained an expression that determines the parameters values for which the hyperbolic points of the main resonance move down to $I=0$. When the hyperbolic points are located on $I=0$, the islands of the main resonance achieve their maximum size \cite{deSousa2010}. Therefore, the initial energy of the particles is close to their rest energy, whereas their final energy is maximum. This is the optimum condition for particle acceleration from low initial energies.

Nonetheless, the condition for optimum acceleration can only be achieved for a limited interval in the wave amplitude $\varepsilon$. For small values of $\varepsilon$ as in Fig. \ref{fig:OriginalSystemRes11}.(a), the hyperbolic points of the main resonance are not located on the axis $I=0$, the islands present a small size, and the acceleration process is not efficient. Increasing the value of $\varepsilon$, the hyperbolic points move down to $I=0$, and the islands size increases. At some point, the system becomes chaotic for high wave amplitudes, and the islands of the main resonance are destroyed by chaos, as can be seen in Fig. \ref{fig:OriginalSystemRes11}.(b). The islands that remain in this figure are too distorted and they are not suitable for regular particle acceleration.

%---------------------------------------------------------------------------------
\section{Improving particle acceleration}
\label{Sec:RobustTorus}

The purpose of our work is to overcome the problems described in the previous section and achieve the condition for optimum acceleration. To do so, we should reduce the initial energy of the particles to their rest energy, and control chaos in the system to prevent it from destroying resonant islands. In this paper, we discuss a method to accomplish that and improve particle acceleration: the addition of a perturbing invariant robust barrier to the system.

The perturbing barrier is a robust invariant curve in phase space generated by an external perturbation. The barrier does not alter the main structures of phase space, but it reduces the perturbation caused by the wave and controls chaos in a region around it \cite{EgydiodeCarvalho2009, Martins2010, Martins2011}. On the other hand, the barrier amplifies the action of the wave on the particles for regions far from it in phase space.

Such perturbing robust barriers have been investigated theoretically for fusion devices. References \cite{EgydiodeCarvalho2009, Martins2010, Martins2011} show that the barrier controls the transport, improves plasma confinement, and prevents the plasma from reaching and damaging the tokamak walls. The presence of transport barriers were also observed experimentally in tokamaks. In Ref. \cite{Marcus2008}, the barrier was created by alterations in the plasma electric field produced by electrodes placed at the plasma edge. References \cite{Frerichs2012, FraileJunior2017} describe the plasma response to a resonant magnetic perturbation generated by an external current, which produces effects similar to a robust barrier.

For wave-particle interactions, the creation of multiple barriers in phase space has already been achieved experimentally \cite{Chandre2005} with the purpose of controlling chaos in the particles dynamics. In this paper, we consider a single perturbing robust barrier in phase space, and we investigate theoretically how this robust barrier can be used to improve particle acceleration in plasma based accelerators.

For low density beams, wave-particle interactions are described by Hamiltonian (\ref{eq:HOriginalCart}). The exact physical configuration of the barrier depends on the external perturbation used to generate it in the experiment. However, the addition of a typical perturbing robust barrier to the system, as the ones observed in the experiments described by Refs. \cite{Marcus2008, Frerichs2012, FraileJunior2017}, produces a Hamiltonian on the form of (\ref{eq:HRobustTorus}) for wave-particle interactions:
\begin{align}
	\label{eq:HRobustTorus}
	H = \sqrt{1 + 2I} + & \frac{\varepsilon}{2} (I-I_{\text{PRB}})^2 \cos(k\sqrt{2I}\sin\theta) \nonumber \\*
						& \times \displaystyle\sum_{n=-\infty}^{+\infty} \delta(t-nT) ,
\end{align}
where $I_{\text{PRB}}$ indicates the position of the perturbing robust barrier in phase space. For trajectories with $I$ close to $I_{\text{PRB}}$, the effective amplitude of the wave is reduced, and for $I=I_{\text{PRB}}$ the perturbation vanishes.

The strategy we used in Section \ref{Sec:OriginalSystem} allowed us to obtain an exact and explicit map for the original system. This strategy could be applied to the original system because the perturbative term in Hamiltonian (\ref{eq:HOriginalCart}), as well as the integrable term in Hamiltonian (\ref{eq:HOriginalITheta}), depends only on one of the two canonical variables. When we add a perturbing robust barrier to the system, the Hamiltonian in Cartesian coordinates equivalent to Hamiltonian (\ref{eq:HRobustTorus}) does not have this property. Thus, the strategy we used for the original system cannot be applied to the perturbed system.

Instead, we integrate Hamiltonian (\ref{eq:HRobustTorus}) assuming that $\theta \rightarrow \theta_n$ and $I \rightarrow I_{n+1}$ across a wave pulse:
\begin{subequations}
	\label{eq:TEAcrossPulseItheta}
	\begin{align}
		I_n^+ = & I_n + \frac{\varepsilon k}{2} (I_{n+1} - I_{\text{PRB}})^2 \sqrt{2I_{n+1}}   \nonumber \\*
				& \times \cos\theta_n \sin(k \sqrt{2I_{n+1}} \sin\theta_n) , \\*
		\theta_n^+ = & \theta_n - \frac{\varepsilon k}{2} \frac{(I_{n+1} - I_{\text{PRB}})^2}{\sqrt{2I_{n+1}}} \sin\theta_n \sin(k \sqrt{2I_{n+1}} \sin\theta_n)   \nonumber \\*
					 & + \varepsilon (I_{n+1} - I_{\text{PRB}}) \cos(k \sqrt{2I_{n+1}} \sin\theta_n) .
	\end{align}
\end{subequations}

Between two consecutive wave pulses, the perturbed Hamiltonian (\ref{eq:HRobustTorus}) is integrable and equal to the original Hamiltonian (\ref{eq:HOriginalITheta}). Thus, the changes in the $(I,\theta)$ variables between two pulses for the perturbed system are also given by expressions (\ref{eq:TEBetweenPulsesITheta}).

Using expressions (\ref{eq:TEAcrossPulseItheta}) and (\ref{eq:TEBetweenPulsesITheta}), we obtain a map that describes the time evolution of the perturbed system:
\begin{figure*}[!tb]
	\centering
	\includegraphics{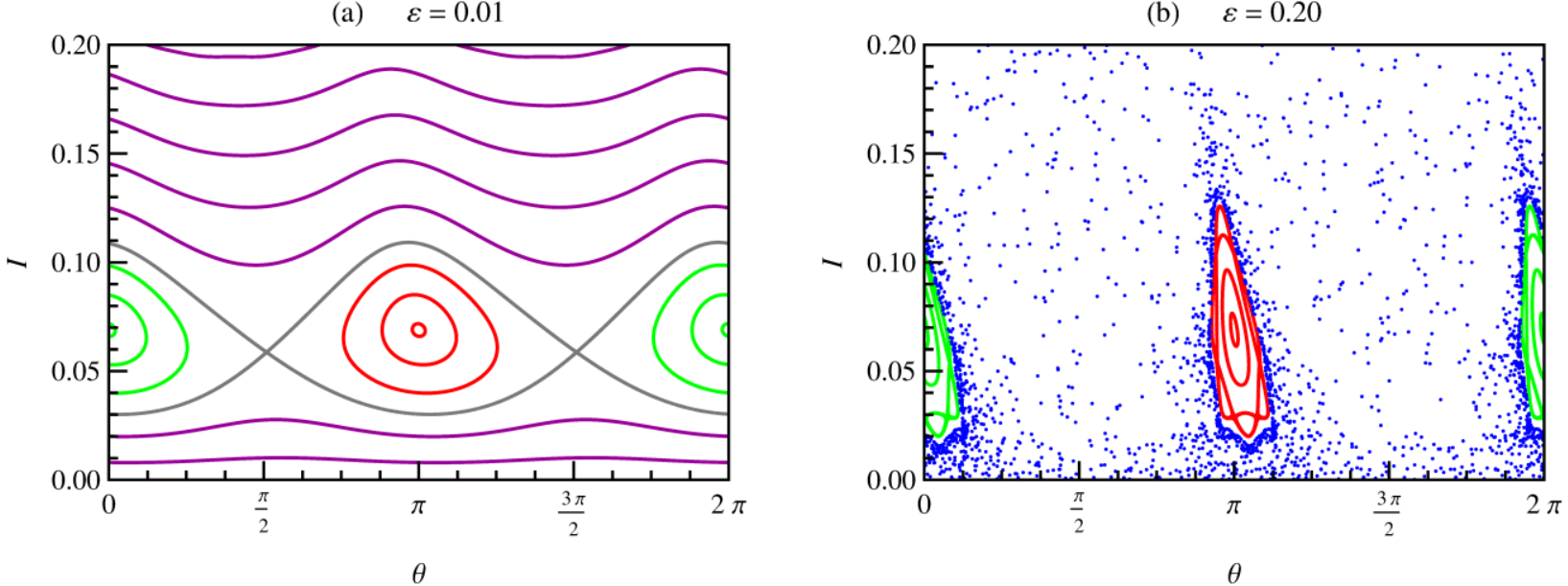}
	\caption{Amplification of Fig. \ref{fig:OriginalSystem} in the main resonance region.}
	\label{fig:OriginalSystemRes11}
\end{figure*}
\begin{widetext}
	\begin{subequations}
		\label{eq:MapRobustTorus}
		\begin{align}
			I_{n+1} = & I_n + \frac{\varepsilon k}{2} (I_{n+1} - I_{\text{PRB}})^2 \sqrt{2I_{n+1}} \cos\theta_n \sin(k \sqrt{2I_{n+1}} \sin\theta_n) , \\*
			\theta_{n+1} = & \theta_n - \frac{\varepsilon k}{2} \frac{(I_{n+1} - I_{\text{PRB}})^2}{\sqrt{2I_{n+1}}} \sin\theta_n \sin(k \sqrt{2I_{n+1}} \sin\theta_n) + \varepsilon (I_{n+1} - I_{\text{PRB}}) \cos(k \sqrt{2I_{n+1}} \sin\theta_n)    \nonumber \\*
						   & + \frac{T}{\sqrt{1 + 2I_{n+1}}} \pmod{2\pi} .
		\end{align}
	\end{subequations}
\end{widetext}

The assumption that $\theta \rightarrow \theta_n$ and $I \rightarrow I_{n+1}$ across a wave pulse is necessary to guarantee the symplectic character of the Hamiltonian system. Nonetheless, the map we obtain is implicit in the action variable $I$. Unlike map (\ref{eq:MapOriginal}), map (\ref{eq:MapRobustTorus}) for the perturbed system is approximate, linear in $\varepsilon$, and valid only for $\varepsilon I \ll 1$.

From map (\ref{eq:MapRobustTorus}), we determine a relation for the parameters of the system that brings the periodic points of the main resonance down to the axis $I=0$. For these periodic points, $\theta_{n+1} = \theta_n + 2\pi$. We replace this result in the second equation of map (\ref{eq:MapRobustTorus}), and then we calculate the limit $I_{n+1} \rightarrow 0$, obtaining the following expression
\begin{equation}
	\label{eq:I=0}
	\sin\theta_n = \pm \sqrt{\frac{2}{\varepsilon k^2 I_{\text{PRB}}^2} (T - \varepsilon I_{\text{PRB}} - 2\pi)} .
\end{equation}

For the hyperbolic points of the main resonance, $\theta_n = \pi/2; \, 3\pi/2$, $\sin\theta_n = \pm 1$, and expression (\ref{eq:I=0}) yields
\begin{equation}
	\label{eq:I=0HyperbolicPts}
	(I_{\text{PRB}})_{\text{hp}} = \frac{-\varepsilon + \sqrt{\varepsilon^2 - 2\varepsilon k^2 (2\pi - T)}}{\varepsilon k^2} .
\end{equation}
Whenever $I_{\text{PRB}} \geq (I_{\text{PRB}})_{\text{hp}}$, the hyperbolic points of the main resonance are located over the axis $I=0$, meaning that it is possible to accelerate particles from their rest energy.

It is important to notice that the periodic points in $\theta_n = \pi/2; \, 3\pi/2$ are hyperbolic for \cite{deSousa2013}
\begin{equation}
	\label{eq:IsochronousBifurcation}
	k < (k)_{\text{ib}} = \frac{\pi}{\sqrt{2I_{1,1}}} ,
\end{equation}
where $I_{1,1}$ is the position of the elliptic points in $\theta_n=0; \, \pi$ with respect to the action variable. When $k = (k)_{\text{ib}}$, the hyperbolic points undergo an isochronous bifurcation and become elliptic, as we describe in Ref. \cite{deSousa2013}. Therefore, condition (\ref{eq:I=0HyperbolicPts}) is only valid for values of $k$ that satisfy equation (\ref{eq:IsochronousBifurcation}).

In the main resonance, the elliptic points are located on $\theta_n=0; \, \pi$, such that $\sin\theta_n = 0$, and from expression (\ref{eq:I=0}) we obtain
\begin{equation}
	\label{eq:I=0EllipticPts}
	(I_{\text{PRB}})_{\text{ep}} = \frac{T - 2\pi}{\varepsilon} .
\end{equation}
When $I_{\text{PRB}} = (I_{\text{PRB}})_{\text{ep}}$, the elliptic points touch the axis $I=0$. For $I_{\text{PRB}} > (I_{\text{PRB}})_{\text{ep}}$, the resonant islands are no longer present in phase space. Therefore, to accelerate particles in the main resonance of the system, it is necessary that $I_{\text{PRB}} < (I_{\text{PRB}})_{\text{ep}}$.

Besides condition (\ref{eq:I=0EllipticPts}) that brings the elliptic points down to the axis $I=0$, the islands of the main resonance also disappear from phase space when the elliptic points undergo a period doubling bifurcation. In this type of bifurcation, the elliptic points lose stability, becoming hyperbolic. Each primary resonant island is replaced by a pair of smaller secondary islands that are not suitable for particle acceleration.

The periodic points of the main resonance located on $\theta=0; \, \pi$ are elliptic as long as the eigenvalues of the Jacobian matrix calculated in these periodic points are complex conjugates. It means that the periodic points on $\theta=0; \, \pi$ are elliptic for values of $I_{\text{PRB}}$ in the interval
\begin{subequations}
	\label{eq:PDBifurcation}
	\begin{align}
		I_{\text{PRB}} > & (I_{\text{PRB}})_{\text{pdb}} = I_{1,1} \nonumber \\*
						 & - 2\left[\varepsilon k^2 I_{1,1} \left(\frac{T}{(1 + 2I_{1,1})^{3/2}} - \varepsilon \right)\right]^{-1/2} , \\*
		I_{\text{PRB}} < & (I_{\text{PRB}})_{\text{pdb}} = I_{1,1} \nonumber \\*
						 & + 2\left[\varepsilon k^2 I_{1,1} \left(\frac{T}{(1 + 2I_{1,1})^{3/2}} - \varepsilon \right)\right]^{-1/2} ,
	\end{align}
\end{subequations}
with
\begin{subequations}
	\label{eq:PDBifurcationCond}
	\begin{align}
		&I_{\text{PRB}} \geq 0 , \\*
		&\frac{\varepsilon}{T} < \frac{1}{(1 + 2I_{1,1})^{3/2}} .
	\end{align}
\end{subequations}

In a first order approximation, the position $I_{1,1}$ of the elliptic points can be calculated analytically as \cite{deSousa2013}
\begin{equation}
	\label{eq:I1,1Aprox}
	I_{1,1} \approx \frac{T^2}{8\pi^2} - \frac{1}{2} .
\end{equation}
Expression (\ref{eq:I1,1Aprox}) implies that the islands of the main resonance are only present in phase space for $T>2\pi$, since the action $I$ is a positive quantity.

Replacing the value of $I_{1,1}$ from expression (\ref{eq:I1,1Aprox}) in (\ref{eq:PDBifurcation}), we have
\begin{subequations}
	\label{eq:PDBifurcationApprox}
	\begin{align}
		I_{\text{PRB}} > & (I_{\text{PRB}})_{\text{pdb}} = \frac{T^2}{8\pi^2} - \frac{1}{2} \nonumber \\*
						 & - 2\left[\varepsilon k^2 \left(\frac{T^2}{8\pi^2} - \frac{1}{2}\right) \left(\frac{8\pi^3}{T^2} - \varepsilon \right)\right]^{-1/2} , \\*
		I_{\text{PRB}} < & (I_{\text{PRB}})_{\text{pdb}} = \frac{T^2}{8\pi^2} - \frac{1}{2} \nonumber \\*
						 & + 2\left[\varepsilon k^2 \left(\frac{T^2}{8\pi^2} - \frac{1}{2}\right) \left(\frac{8\pi^3}{T^2} - \varepsilon \right)\right]^{-1/2} ,
	\end{align}
\end{subequations}
whereas condition (\ref{eq:PDBifurcationCond}) becomes
\begin{subequations}
	\label{eq:PDBifurcationCondAprox}
	\begin{align}
		&I_{\text{PRB}} \geq 0 , \\*
		&\varepsilon T^2 < 8\pi^3 .
	\end{align}
\end{subequations}
\begin{figure}[!tb]
	\centering
	\includegraphics{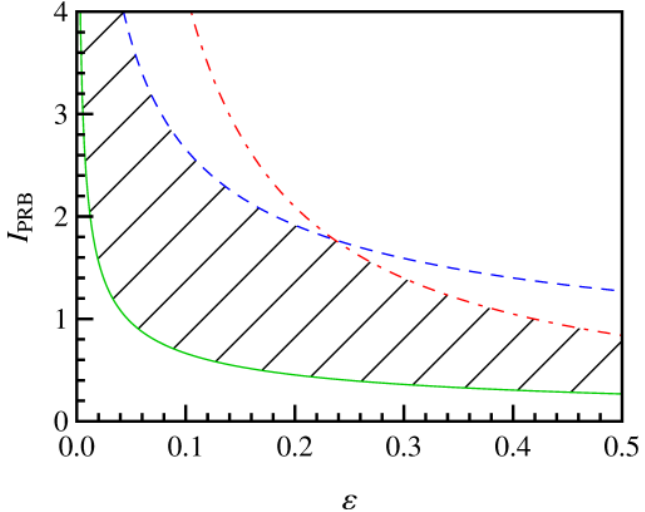}
	\caption{(Color online) Above the solid green curve, the hyperbolic points in $\theta = \pi/2; \, 3\pi/2$ are located on the axis $I=0$. On the dot-dashed red curve, the elliptic points in $\theta=0; \, \pi$ move down to $I=0$. On the dashed blue curve, the elliptic points in $\theta=0; \, \pi$ undergo a period doubling bifurcation. Thus, regular particle acceleration from their rest energy is only possible for values of $\varepsilon$ and $I_{\text{PRB}}$ within the hatched area. Other parameters in this figure are $T=2 \pi (1 + 1/15)$, and $k=4$.}
	\label{fig:ParametersAcceleration}
\end{figure}

In Fig. \ref{fig:ParametersAcceleration}, we represent expressions (\ref{eq:I=0HyperbolicPts}), (\ref{eq:I=0EllipticPts}), and (\ref{eq:PDBifurcationApprox}) in a graphic of $I_{\text{PRB}} \times \varepsilon$ for $T=2 \pi (1 + 1/15)$ and $k=4$. According to expression (\ref{eq:I=0HyperbolicPts}), the solid green curve represents the values of $I_{\text{PRB}}$ for which the hyperbolic points in $\theta = \pi/2; \, 3\pi/2$ reach the axis $I=0$. Thus, to accelerate particles from their rest energy, the position $I_{\text{PRB}}$ of the perturbing barrier must be above this curve.
\begin{figure*}[!tb]
	\centering
	\includegraphics{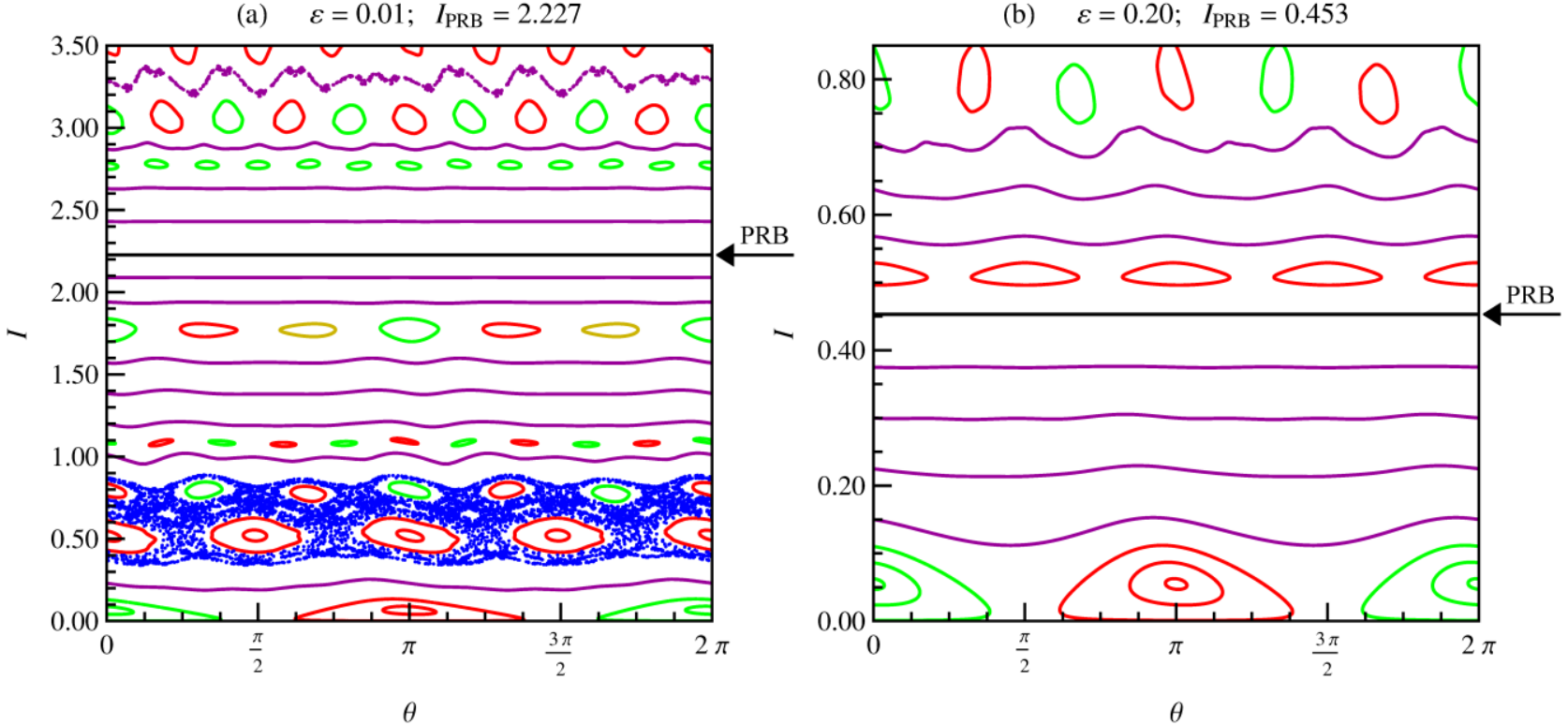}
	\caption{Phase space of the perturbed system for $T=2 \pi (1 + 1/15)$, $k=4$, (a) $\varepsilon=0.01$, and $I_{\text{PRB}}=2.227$; (b) $\varepsilon=0.20$, and $I_{\text{PRB}}=0.453$. The arrows indicate the position of the perturbing robust barrier (PRB).}
	\label{fig:RobustTorus}
\end{figure*}

The dot-dashed red curve was built from expression (\ref{eq:I=0EllipticPts}). Over the curve, the elliptic points on $\theta=0; \, \pi$ are located on the axis $I=0$. Above the curve, the islands of the main resonance disappear from phase space. The dashed blue curve represents expressions (\ref{eq:PDBifurcationApprox}) for which the periodic points on $\theta=0; \, \pi$ undergo a period doubling bifurcation. Under the curve, these points are elliptic. Above the curve, they are hyperbolic and the primary resonant islands no longer exist.

Therefore, particle acceleration in the islands of the main resonance is only possible under the dot-dashed red and the dashed blue curves. Furthermore, if one wants to accelerate particles from their rest energy, the values of $\varepsilon$ and $I_{\text{PRB}}$ must be above the solid green curve and inside the hatched area in Fig. \ref{fig:ParametersAcceleration}.

%---------------------------------------------------------------------------------
\section{Results and discussion}
\label{Sec:ResultsDiscussion}

In the original system, only the hyperbolic points move down to $I=0$, whereas the elliptic points remain essentially in the same position \cite{deSousa2010}. When we add a perturbing robust barrier to the system, the elliptic points also move down, reducing the islands size and, thus, the final energy of the particles. To keep the final energy to a maximum, the position $I_{\text{PRB}}$ of the perturbing barrier should be as low as possible. The best position for the perturbing barrier is exactly over the solid green curve in Fig. \ref{fig:ParametersAcceleration}, i.e. $I_{\text{PRB}} = (I_{\text{PRB}})_{\text{hp}}$, such that the initial energy of the particles is close to their rest energy and their final energy is maximum.

In Fig. \ref{fig:RobustTorus}, we show the phase space of the perturbed system for the same parameters used in Fig. \ref{fig:OriginalSystem}. The arrows in the figures indicate the position $I_{\text{PRB}}$ of the perturbing robust barrier. We choose $I_{\text{PRB}} = (I_{\text{PRB}})_{\text{hp}}$ to obtain the optimum condition for particle acceleration.

It is important to notice that the perturbing robust barrier alters the action of the wave on the particles for the whole phase space. In the original system, the amplitude of the wave is constant and equal to $\varepsilon/2$. With the perturbing barrier, the wave presents an effective amplitude $\varepsilon (I-I_{\text{PRB}})^2 /2$ that varies in phase space. For regions with $(I-I_{\text{PRB}}) > 1$, the effective wave amplitude is higher than the original wave amplitude $\varepsilon/2$, which increases the resonant islands size in such regions of phase space. On the other hand, when $(I-I_{\text{PRB}}) < 1$, the effective wave amplitude is lower than the original wave amplitude, and the chaotic regions are reduced in phase space.

In Fig. \ref{fig:RobustTorus}.(a), the perturbing barrier is placed far from the main resonance. The effective wave amplitude is higher than the original wave amplitude in the region $I<1.0$, and it increases the islands size in this region when compared to Fig. \ref{fig:OriginalSystem}.(a). The presence of the barrier also produces chaotic trajectories around the resonant islands in the region $0.35<I<0.90$. Since the effective wave amplitude is small, the chaotic trajectories are confined and they do not reach the islands of the main resonance, which are located in a stable region.

Increasing the wave amplitude to $\varepsilon=0.20$, the phase space of the original system in Fig. \ref{fig:OriginalSystem}.(b) is covered by a chaotic sea that distorts and destroys the islands of the main resonance. When we add a perturbing barrier as in Fig. \ref{fig:RobustTorus}.(b), it reduces the effective wave amplitude in the region $I<1.453$, thus controlling chaos in the system. The perturbing barrier regularizes the phase space and restores the islands of the main resonance, which can be used for particle acceleration once more.

\begin{figure*}[!tb]
	\centering
	\includegraphics{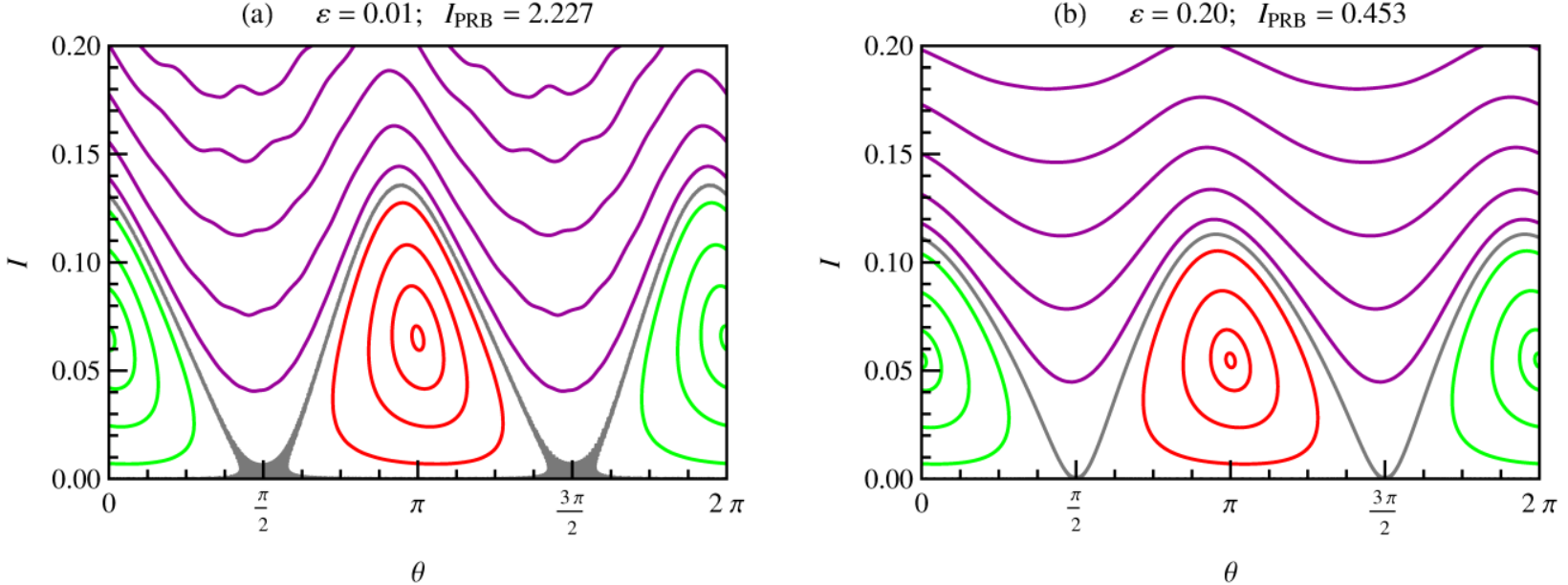}
	\caption{Amplification of Fig. \ref{fig:RobustTorus} in the main resonance region.}
	\label{fig:RobustTorusRes11}
\end{figure*}

Figure \ref{fig:RobustTorusRes11} shows an amplification of Fig. \ref{fig:RobustTorus} in the main resonance region. In Fig. \ref{fig:OriginalSystemRes11}.(a), the resonant islands of the original system present a usual shape, with the elliptic and hyperbolic points approximately in the same position with respect to the action variable. The separatrix does not touch the axis $I=0$, and it is not possible to accelerate particles from their rest energy. In Fig. \ref{fig:RobustTorusRes11}.(a), the perturbing barrier increases the effective wave amplitude in the main resonance region, and the islands of the perturbed system acquire a triangular shape. The perturbing barrier brings the hyperbolic points down to the axis $I=0$, and the initial energy of the particles is close to their rest energy.

We obtain the dimensionless momentum from the canonical transformation $p_x = \sqrt{2I}\cos\theta$, whereas the dimensionless velocity is given by \cite{deSousa2010} $(v_x)_{\text{dl}} = p_x (1 + p_x^2)^{-1/2}$. The one-dimensional velocity in the $x$ direction is calculated in units of $c$ as $v_x = c(v_x)_{\text{dl}}$, with $c$ the speed of light. In the original system, for $\varepsilon=0.01$ the initial velocity of the particles is $v_{x,i} \simeq 0.239c$, and their final velocity is $v_{x,f} \simeq 0.423c$. In the perturbed system, the initial velocity is $v_{x,i} \simeq 0.064c$, and the final velocity is $v_{x,f} \simeq 0.453c$. It means that the perturbing barrier reduces the initial velocity of the particles in approximately $73\%$, and it increases their final velocity is approximately $7\%$.

In Fig. \ref{fig:OriginalSystemRes11}.(b), $\varepsilon=0.20$ and the hyperbolic points of the main resonance are located over the axis $I=0$. However, the external trajectories of the resonant islands were destroyed by chaos, and the internal trajectories that remain are too distorted and not suitable for particle acceleration. In this scenario, the initial velocity of the particles is $v_{x,i} \simeq 0.235c$, whereas their final velocity is $v_{x,f} \simeq 0.414c$.

When we add a perturbing robust barrier to the system as in Fig. \ref{fig:RobustTorusRes11}.(b), it reduces the effective wave amplitude in the main resonance region, controlling chaos and restoring regular trajectories. The barrier recovers the islands of the main resonance, and it also brings the hyperbolic points down to the axis $I=0$. For $\varepsilon = 0.20$, the initial velocity of the particles is $v_{x,i} \simeq 0.0046c$, and their final velocity is $v_{x,f} \simeq 0.427c$. In the perturbed system, the initial velocity of the particles is approximately $98\%$ lower than in the original system, and their final velocity is approximately $3\%$ higher.

From these results, we conclude that a properly located perturbing robust barrier is a very efficient method to improve particle acceleration. The perturbing barrier increases the wave amplitude interval for which we can achieve the optimum condition for particle acceleration. For low amplitude waves as in Fig. \ref{fig:RobustTorusRes11}.(a), the barrier drastically reduces the initial velocity we should provide to the particles in the beginning of the acceleration process, and it slightly increases the final velocity of particles. With the perturbing barrier, the final velocity reaches its maximum values for low amplitude waves, meaning that it is possible to accelerate particles to high final velocities even for low amplitude waves that require less energy to be produced.

For high amplitude waves as in Fig. \ref{fig:RobustTorusRes11}.(b), the perturbing barrier controls chaos in the system, and restores the acceleration process. However, the final velocity of particles decreases with the wave amplitude in the perturbed system. If we continue raising the wave amplitude, the acceleration process will no longer be effective. According to expression (\ref{eq:I=0EllipticPts}), for sufficiently high wave amplitudes, the perturbing barrier brings the elliptic points of the resonance down to $I=0$, and the islands used for particle acceleration disappear from phase space.

In the main resonance of the system, the final velocity reached by the particles is lower than $0.5c$. Even though we concentrate our analysis in this region of phase space, the use of a relativistic Hamiltonian is necessary to properly reproduce all the dynamical features of the system. If the relativistic effects are ignored, the islands of the main resonance are replaced by invariant tori in which the particles cannot be accelerated by the wave. For wave-particle interactions as the one we analyze in this paper, the role of nonlinearities arising from relativistic mass correction is determinant to the dynamical characteristics of the system. Furthermore, the perturbing robust barrier may be placed in regions of relativistic velocities, such as in Figs. \ref{fig:RobustTorus}.(a) and \ref{fig:RobustTorusRes11}.(a), where the perturbing barrier is in a region of velocities on the order of $0.9c$.

%---------------------------------------------------------------------------------
\section{Conclusions}
\label{Sec:Conclusions}

We analyzed a relativistic low density beam confined by a uniform magnetic field in the axial direction, and perturbed by a stationary electrostatic wave propagating perpendicularly to the magnetic field. In this system, resonant islands are used for particle acceleration since the wave transfers a great amount of energy to the particles in these trajectories. In a previous work, we demonstrated that, according to the parameters of the wave, the islands of the main resonance may present a triangular shape, with the hyperbolic points located over the axis $I=0$, which corresponds to the rest energy of the particles. When the hyperbolic points are located over $I=0$, the initial energy of the particles is minimum and it is close to their rest energy. Moreover, the islands reach their maximum size, and the final energy of the particles is maximum. This is the optimum condition for particle acceleration from low initial energies in plasmas.

In this paper, we showed that the optimum condition for particle acceleration is not reached for small values of the wave amplitude. In this case, the islands of the main resonance are not triangular shaped, and it is not possible to accelerate particles from their rest energy. Increasing the wave amplitude, the hyperbolic points move down to $I=0$. However, the system becomes chaotic for high values of the wave amplitude. Resonant islands are partially destroyed and what remains is too distorted and not suitable for particle acceleration.

To overcome these problems, we introduced a perturbing robust barrier in the system. An external perturbation generates a robust invariant barrier that preserves the main structures of phase space. The barrier reduces the perturbation caused by the wave around it, and at the same time, it amplifies the action of the wave on particles for regions far from it in phase space.

For the perturbed system, we obtained the position of the robust barrier that brings the hyperbolic points down to the axis $I=0$. We also calculated the position of the barrier for which the resonant islands are no longer present in phase space, either because the elliptic points move down to $I=0$, or because they lose stability through a period doubling bifurcation. From these results, we obtained an interval for the position of the perturbing barrier that allows us to accelerate particles from their rest energy in the main resonance of the system.

We also determined the best position for the perturbing barrier inside this interval. We showed that, with the barrier, it is possible to achieve the optimum condition for particle acceleration for a much larger interval in the wave amplitude. For low values of the wave amplitude, the perturbing barrier brings the hyperbolic points down to $I=0$, and particles can be accelerated from their rest energy. When the wave amplitude is high, the perturbation controls chaos in the system and restores the islands of the main resonance used for particle acceleration from rest energy. In both cases, the perturbation slightly increases the final energy of the particles.

From these results, we conclude that the perturbing robust barrier is an efficient method to improve particle beam acceleration in plasma based accelerators. It reduces the initial energy of the particles to their rest energy, increases the final energy of the particles, and controls chaos around it in phase space.

In this paper, we focused our analysis in the main resonance of the system, for which the initial energy of particles is close to their rest energy. Nonetheless, the perturbing robust barrier can be used likewise to improve particle acceleration in the other resonances the system presents, including resonances where the velocity of the particles is close to the speed of light. We also point out that the procedure we presented, use of a perturbing barrier to control chaos and improve particle acceleration, could similarly be applied to other systems to improve applications of interest. In the case of fusion devices for example, the perturbing barrier is useful to control particle transport and prevent the plasma from reaching and damaging the tokamak walls.

%---------------------------------------------------------------------------------
\begin{acknowledgments}
We acknowledge financial support from the Brazilian scientific agencies: S\~ao Paulo Research Foundation (FAPESP) under Grants No. 2015/05186-0 and No. 2011/19296-1, Conselho Nacional de Desenvolvimento Cient\'ifico e Tecnol\'ogico (CNPq), and Coordena\c{c}\~ao de Aperfei\c{c}oamento de Pessoal de N\'ivel Superior (Capes).
\end{acknowledgments}

%---------------------------------------------------------------------------------
\bibliography{RobustTorusBib}

\end{document}